\documentclass[english]{article}
\usepackage[T1]{fontenc}
\usepackage[latin9]{inputenc}
\usepackage{amssymb}
\usepackage{graphicx}
\usepackage{babel}
\begin{document}

\title{Pre-big bang scenario and the WZW model}

\author{Marcel Jacon}
\maketitle
\begin{abstract}
Extensive studies of pre-big bang scenarios for Bianchi-I type universe
have been made, at various approximation levels. Knowing the solution
of the equations for the post-big bang universe, the symmetries of
the equations (``time reversal and scale dual transformations'')
allow the study of pre-big bang solutions. However, they exhibit singularitites
in both the curvature and the dilaton kinetic energy. 

Calculating the $\beta$ equations for the Non-Linear Sigma model,
at the first loop approximation and imposing conformal invariance
at this level, lead to equations of motion that simply state that
the curvature must be nil , which in turn allows the utilization of
groups to solve the $\beta$ equations. This is what is done in the
Weiss-Zumino-Witten (WZW) model. 

In this article, we will show that using the WZW model on $SU_{2}$
,some of the difficulties encountered in the determination of the
pre and post big-bang solutions are eliminated, leading to realistic
solutions for the evolution of the universe and giving an explanation
to the actually observed acceleration of the expansion. 
\end{abstract}

\section{Introduction:}

Extensive studies of pre-big bang scenarios for Bianchi-I type universe
have been made at various approximation levels {[}1,2{]} and references
herein. The gravitational (massless, bosonic) sector of the string
action contains not only the metric, but also ,at least one more fondamental
field, the dilation $\phi$.The corresponding tree-level action lead
to cosmological equations which have been established in the case
where the (NS-NS) two form $B{\scriptstyle \mu{\scriptstyle \nu={\textstyle 0}}},$
but including the contribution of perfect fluid sources.These equations
are invarariant under ``time reversal transformation, but also under
``scale dual transformations''. Knowing the solution of the equations
for the post-big bang universe, these symmetry allow the study of
pre-big bang solutions. The exact integration of the string cosmology
equations in the fully anisotropic case can be performed but lead
to solutions which exhibit singularitites in both the curvature and
the dilaton kinetic energy. The solutions associated with the pre
and post-big bang branches, being disconnected by a singularity, are
not appropriate to describe the whole transition between the two regimes.
Particular examples of regular solutions may be obtained, but one
may also expect that the regularisation of the big bang singularity
need also to introduce the effects of higher order loop and $\alpha{\scriptscriptstyle '}$
corrections. 

The Polyakov action may be modified to incorporate the effect of massless
excitation, and leads to theNon-Linear Sigma model {[}3,4{]}. Calculating
the $\beta$ equations for the Non-Linear Sigma (NLS) model at the
first loop approximation and imposing conformal invariance at this
level, leads to equations of motion that simply state that the curvature
must be nil, which in turn allows the utilization of group manifolds
to solve the $\beta$ equations. This is what is done in the Weiss-Zumino-Witten
(WZW) model {[}5{]}

In this article, we will show how, using the WZW model on $SU_{2}$
,some of the difficulties encountered in the determination of the
pre and post big-bang solutions are eliminated, leading to realiistic
solutions for the evolution of the universe.

\section{The WZW-model on $SU_{2}$.}

An element of $SU_{2}$ is parametrized by means of the well-known
Euler angles $(\alpha,\beta,\gamma)$:

\begin{center}
$g=\exp(\alpha\tau_{3})$$\exp(\beta\tau_{2})\exp(\gamma\tau_{3})$
\par\end{center}

where the $\tau_{i}$ matrices are given in terms of Pauli matrices
$\tau_{k}$=$\frac{\sigma_{k}}{2i}$, and the Euler angles have values
in the ranges 0$\leq$$\alpha$<2$\pi$, 0$\leq\beta$<$\pi$, 0$\leq\gamma$<2$\pi$

Using the expressions for the Maurer-Cartan forms {[}6{]} , one deduces
the line element:

ds$^{2}$=d$\alpha^{2}$+d$\beta^{2}$+d$\gamma^{2}$+2d$\alpha$d$\gamma$$\cos\beta$
which determines the $g$ matrix 

\begin{center}
$g$=$\left(\begin{array}{ccc}
1 & 0 & \cos\beta\\
0 & 1 & 0\\
\cos\beta & 0 & 1
\end{array}\right)$
\par\end{center}

\begin{flushright}
(2.0.1)
\par\end{flushright}

whose invere is 

\begin{center}
$g^{-1}$=$\frac{1}{\sin^{2}\beta}\left(\begin{array}{ccc}
1 & 0 & -\cos\beta\\
0 & \sin^{2}\beta & 0\\
-\cos\beta & 0 & 1
\end{array}\right)$
\par\end{center}

\begin{flushright}
(2.0.2)
\par\end{flushright}

From the elements of these matrix, one easily obtain the Christoffel
symbols and finally the Ricci tensors elements and the curvature scalar

\begin{center}
\textit{$R$=$R_{\mu\nu}$g$^{\mu\nu}$=$\frac{3}{2}$}
\par\end{center}

\begin{flushright}
(2.0.3)
\par\end{flushright}

A space of constant curvature is a conformally flat space, so that
its Weyl tensor vanishes identically.

\subsection{The evolution of the universe:}

When studying the evolution on the universe, we work in the \textit{S-frame
representation }of the string effective action\cite{key-2}. Assuming
that $B_{\mu\nu}$ is vanishing and that the matter sources can be
represented by a fluid, in the synchronous gauge of the comoving frame,
we have:

\begin{center}
$\tilde{g}_{\mu\nu}=diag(1,$-$\tilde{a}^{2}$$\delta_{ij})$, $\tilde{a}=$$\tilde{a}(t)$,
$\tilde{\phi}=\tilde{\phi}(t)$
\par\end{center}

\begin{center}
$\tilde{T_{\mu}^{\nu}}$=diag($\tilde{\rho}-\tilde{p}\delta_{ij})$,
$\tilde{\rho}=\tilde{\rho}(t)$, $\tilde{p}=\tilde{p}(t),$ $\tilde{\sigma}=\tilde{\sigma}(t)$
\par\end{center}

These definitions and the use of the cosmic time as the time coordinate
leads to the well-known Roberson-Walker metric

\begin{center}
$ds^{2}$=$dt^{2}-\tilde{a}^{2}(t)\left[(dr^{2}/1-Kr^{2})+r^{2}(d\theta^{2}+sin^{2}\vartheta d\phi^{2})\right]$
\par\end{center}

In this formula, $K$ has dimension $L^{-2}$ and $\tilde{a}$(t)
is dimensionless.

In order to preserves the $SU_{2}$ character of the 3-dimensional
spatial subgroup and keep its interesting symmetry properties, such
as a constant curvature ( i.e. independent of the spatial coordinates),
for a fixed $t,$ we will work in the Euler Angles frame which is
described now. 

We look for a coordinate system $(\xi,\eta,\zeta)$ which diagonalizes
the matrix diag (-a$^{2}$,-a$^{2}$,-a$^{2}$)$\times$g

\begin{flushright}
(2.1.1)
\par\end{flushright}

where $a$ depends on $t$ .

One has the following transformation:

\begin{center}
$\left(\begin{array}{ccc}
\frac{1}{\surd2} & 0 & \frac{-1}{\surd2}\\
0 & 1 & 0\\
\frac{1}{\surd2} & 0 & \frac{1}{\surd2}
\end{array}\right)$:$\left(\begin{array}{ccc}
-a^{2} & 0 & -a^{2}\cos\beta\\
0 & -a^{2} & 0\\
-a^{2}\cos\beta & 0 & -a^{2}
\end{array}\right)$$\left(\begin{array}{ccc}
\frac{1}{\surd2} & 0 & \frac{1}{\surd2}\\
0 & 1 & 0\\
\frac{-1}{\surd2} & 0 & \frac{1}{\surd2}
\end{array}\right)$
\par\end{center}

\begin{center}
=$\left(\begin{array}{ccc}
-a^{2}(1-\cos\beta) & 0 & 0\\
0 & -a^{2} & 0\\
0 & 0 & -a^{2}(1+\cos\beta)
\end{array}\right)$
\par\end{center}

From which it follows:

\begin{center}
$-a^{2}(d\alpha^{2}+d\beta^{2}+d\gamma^{2}+2\cos\beta d\alpha d\gamma)=-a^{2}((1-\cos\beta)d\xi^{2}+d\eta^{2}+(1+\cos\beta)d\zeta$$^{2})$
\par\end{center}

\begin{center}
with $\left[d\xi=(d\alpha-d\gamma)/\sqrt{2},d\eta=d\beta,d\zeta=(d\alpha+d\gamma)/\sqrt{2}\right]$
\par\end{center}

The 4-dimensional line element is

\begin{center}
$ $$\left[ds^{2}=dt^{2}-a^{2}((1-\cos\beta)d\xi^{2}+d\eta^{2}+(1+\cos\beta)d\zeta^{2})\right]$
\par\end{center}

from which one sees that the scale factor $a$ has the dimension $L$

In the S-frame, fixing $r=r_{0}$, an arbitrary value, gives $ds^{2}=dt^{2}-$$\tilde{a}^{2}(t)$$\left[r_{0}^{2}(d\theta^{2}+\sin^{2}\theta d\phi^{2})\right]$.

On the other hand, defining $d\theta=d\eta$ and sin$^{2}$$\theta d\phi^{2}=2(\sin^{2}(\beta/2)d\xi^{2}+\cos^{2}(\beta/2)d\zeta^{2})$
gives the expected result $a=\tilde{a}(t)\times$$r_{0}$.

\begin{flushright}
(2.1.2)
\par\end{flushright}

In this coordinate system, we will see that the curvature scalar is
given by the following constant, generalizing (2.0.3): 

\begin{center}
\textit{R=-(3/$2$a$^{2}$)} 
\par\end{center}

\begin{flushright}
(2.1.3)
\par\end{flushright}

As a consequence, the $\beta$-equations for a fixed $t$ at the first
loop in string perturbation theory, imposing that the theory is conformal
invariant at this order, are satisfied .

\begin{flushright}
(2.1.4)
\par\end{flushright}

We study the graviton-dilaton system , setting B$_{\mu\nu}=0$, but
including perfect fluid sources. Therefore $\phi$ =$\phi$(t) ,T$_{\mu}^{\nu}=$diag
($\rho,-p_{i}\delta_{i}^{j}),$ $\rho$=$\rho(t)$,$p_{i}$$=p_{i}$$(t)$,
$\sigma$ =$\sigma(t)$, where

T$_{\mu\nu}$represente the tenseur current density of the matter
sources and$\sigma$ the scalar charge density. 

We first calculate the components of the Christoffell connection $\Gamma$,
and deduce the components of the Ricci tensor:

From

\begin{center}
$\Gamma_{ij}^{k}$=$\frac{1}{2}$$g^{kl}$($\partial_{i}$g$_{jl}$+$\partial_{j}$$g_{il}$-$\partial_{l}$$g_{ij}$)
\par\end{center}

we deduce

\begin{center}
$\Gamma_{0i}^{i}=\frac{\dot{a_{i}}}{a_{i}}=H_{i}$; $\Gamma_{ii}^{0}=\dot{a}_{i}$$a_{i}$;
$\Gamma_{12}^{1}=\Gamma_{21}^{1}=\frac{\partial_{2}a_{1}}{a_{1}}=\frac{1}{2}\frac{a_{3}}{a_{1}}$;
$\Gamma_{23}^{3}=\Gamma_{32}^{3}$=$\frac{\partial_{2}a_{3}}{a_{3}}=-\frac{1}{2}\frac{a_{1}}{a_{3}}$:
$\Gamma_{11}^{2}=-$$\frac{a_{1}}{a_{2}^{2}}$$\partial_{2}a_{1}=-\frac{1}{2}\frac{a_{1}a_{3}}{a_{2}^{2}}$;
$\Gamma_{33}^{2}=-$$\frac{a_{3}}{a_{2}^{2}}$$\partial_{2}a_{3}=\frac{a_{3}a_{1}}{a_{2}^{2}}$
\par\end{center}

One obtains the following expressions, which apart from a constant,
are similar to those calculated by M. Gasperini in the \textit{S-frame}
(see Eqns. 4.10 of \cite{key-2})

\begin{center}
R$_{0}^{0}$=-$\sum_{i}(\dot{H_{i}}$+$H_{i}^{2}$),
\par\end{center}

\begin{center}
R$_{i}^{i}$=-$\dot{H}_{i}$-$H_{i}(\sum_{k}$$H_{k}$)-1/2a$^{2}$
\par\end{center}

\begin{flushright}
(2.1.5)
\par\end{flushright}

The associated scalar curvature is:

\begin{center}
$R$=-$\sum_{i}$(2$\dot{H_{i}}$+$H_{i}^{2}$)-($\sum_{i}$$H$$_{i}$)$^{2}$-3/2$a^{2}$
\par\end{center}

showing the above result (Eqn 2.1.3) for the three dimensional spatial
manifold.

\begin{flushright}
(2.1.6)
\par\end{flushright}

We retain also the following equations for the dilaton field:

\begin{center}
($\nabla\phi)^{2}$=$\dot{\phi^{2}}$, $\nabla^{2}\phi$=$\ddot{\phi}$$+$$\dot{\phi}$$\sum_{i}$$H_{i}$,
$\nabla_{0}$$\nabla^{0}$$\phi$=$\ddot{\phi}$, $\nabla_{i}$$\nabla^{j}$$\phi$=$\dot{\phi}$$H_{i}$$\delta_{i}$$^{j}$
\par\end{center}

We have also the following relations between the quantities in the
\textit{E} and \textit{S} frames:

\begin{center}
$H_{1}$=$H_{2}$=$H_{3}$=$\frac{\dot{a}}{a}$=$\frac{\dot{\tilde{a}}}{\tilde{a}}$=$\tilde{H}$
\par\end{center}

So the equations for the theory are the same, in the S frame or in
the E frame, except the remaining constant in the$R_{i}^{i}$ .

it follows that we can use the results already obtained by M. Gasperini
in the S-frame. 

\begin{flushright}
(2.1.7)
\par\end{flushright}

Setting $2$$\lambda_{s}$$^{d-1}$=1 ,we have the Euler-Lagrange
equation for the dilaton equation (analogous of Eqn 2.24 of \cite{key-2}.

\begin{center}
$\ddot{\phi}$-$\dot{\phi^{2}}$+2$\dot{\phi}$$\sum_{i}$$H_{i}$-$\sum_{i}$(2$\dot{H}_{i}$+$H_{i}^{2}$)-($\sum_{i}$$H_{i}$)$^{2}$-3/2a$^{2}$+$V$-$\frac{\partial V}{\partial\phi}$=$\frac{1}{2}$$e^{\phi}$$\sigma$
\par\end{center}

\begin{flushright}
(2.1.8)
\par\end{flushright}

The ($0,0)$ component of (Eqn.2.24) in \cite{key-2} gives

\begin{center}
$\dot{\phi^{2}}$-2$\dot{\phi}$$\sum_{i}$$H_{i}$+($\sum_{i}$$H_{i}$)$^{2}$-$\sum_{i}($$H_{i}^{2}$)
+(3/2$a^{2}$)-$V$=$e^{\phi}$$\varrho$ 
\par\end{center}

\begin{flushright}
(2.1.9)
\par\end{flushright}

While the diagonal part of the space component gives:

\begin{center}
$\dot{H}_{i}$-$H_{i}$$(\dot{\phi}$-$\sum_{k}$$H_{k}$)+$\frac{1}{2}$$\frac{\partial V}{\partial\phi}$-$(1/4a^{2}$)=$\frac{1}{2}$$e^{\phi}($$p_{i}-\frac{\sigma}{2}$) 
\par\end{center}

\begin{flushright}
(2.1.10)
\par\end{flushright}

These equations are simplified, introducing the ``shifted dilaton
variable `` $\bar{\phi}$ 

\begin{center}
$\bar{\phi}$=$\phi$-$\ln$($a{}^{3}$),$\dot{\bar{\phi}}$=$\dot{\phi}$-$\sum_{i}$$H_{i}$
\par\end{center}

\begin{flushright}
(2.1.11)
\par\end{flushright}

and the shifted variables for the fluid:

\begin{center}
$\bar{\varrho}$=$\rho$($a{}^{3}$), $\bar{p}$=$p$($a{}^{3}$),$\bar{\sigma}$=$\sigma$($a{}^{3}$)
\par\end{center}

then, we obtain the analogous of Eqns 4.39,4.40,4.41 in \cite{key-2}:

\begin{center}
$\dot{\bar{\phi}}$$^{2}$-$\sum_{i}($$H_{i}^{2}$)-$\frac{3}{2a^{2}}$-$V$=$\bar{\rho}$$e^{\bar{\phi}}$
\par\end{center}

\begin{center}
$\dot{H}_{i}$-$H_{i}$$\dot{\bar{\phi}}$+$\frac{1}{2}$$\frac{\partial V}{\partial\bar{\phi}}$-$\frac{1}{4a^{2}}$=$\frac{1}{2}$$e^{\bar{\phi}}$($\bar{p_{i}}$-$\frac{\bar{\sigma}}{2}$)
\par\end{center}

\begin{center}
2$\ddot{\bar{\phi}}$-$\dot{\bar{\phi}}$$^{2}$-$\sum_{i}($$H_{i}^{2}$)-$\frac{3}{2a^{2}}$+$V$-$\frac{\partial V}{\partial\bar{\phi}}$=$\frac{\bar{1}}{2}$$\bar{\sigma}$$e^{\bar{\phi}}$
\par\end{center}

\begin{flushright}
(2.1.12)
\par\end{flushright}

From these equations, we deduce the following conservation equation:

\begin{center}
$\dot{\bar{\rho}}$+$\sum_{i}$$H_{i}$$p_{i}$=$\frac{1}{2}$$\sigma($$\dot{\bar{\phi}}+\sum_{i}$$H_{i})$+$e^{\bar{-\phi}}$$\left[\frac{\partial V}{\partial\bar{\phi}}\sum_{i}H_{i}+\frac{3}{a^{2}}\bar{\phi}-\frac{1}{2a^{2}}\sum_{i}H_{i}-\frac{3}{2}\frac{d}{dt}\frac{1}{a^{2}}\right]$
\par\end{center}

However $V$ is not a scalar under general coordinates transformations,
and it is impossible to define a potential which can be directly inserted
as a scalar into the covariant action. However, it has been shown
that the action and the corresponding equations of motion can be written
in a generalized form which is invariant under general coordinates
transformations using for the potential a non-local variable. 

.The result of the calculations is that the second Eqn. (2.1.12) is
replaced by the simpler one (2.1.13) {[}2,7{]}:

\begin{center}
$\dot{H}_{i}$-$H_{i}$$\dot{\bar{\phi}}$-$\frac{1}{4a^{2}}$=$\frac{1}{2}$$e^{\bar{\phi}}$$\bar{p_{i}}$
\par\end{center}

\begin{flushright}
(2.1.13)
\par\end{flushright}

leading to the modified conservation equation:

\begin{center}
$\dot{\bar{\rho}}$+$\sum_{i}$$H_{i}$$p_{i}$=$\frac{1}{2}$$\sigma$$\dot{\bar{\phi}}$+$e^{\bar{-\phi}}$$\left[^{\dot{\bar{\phi}}}\frac{3}{2a^{2}}-\frac{1}{2a^{2}}\sum_{i}H_{i}-\frac{3}{2}\frac{d}{dt}\frac{1}{a^{2}}\right]$
\par\end{center}

as well as the Eqn for the Dilaton field $\bar{\phi(t):}$ (Eqn 4A.27
of ref.2)

\begin{center}
$\ddot{\phi}$+$\sum_{i}$$H_{i}$$\bar{\phi}$-$\dot{\phi^{2}}$+$2$$\lambda_{s}^{d-1}$$(V-\frac{1}{2}\frac{\partial V}{\partial\phi})$
+$\lambda_{s}^{d-1}$$e^{\phi}$($\rho-$$\sum_{i}$$p_{i}$)+$\frac{1}{2}$($d-1$)$\lambda_{s}^{d-1}$$e^{\phi}$$\sigma$=$0$
\par\end{center}

\begin{flushright}
(2.1.14)
\par\end{flushright}

Using these equations, we can now study models for the evolution of
the Universe.

\subsection{Models for the evolution of the Universe:}

\begin{flushleft}
Instead of introducing a time parameter $x$ as in {[}2{]} and get
a full analytical result, we calculate a numerical solution of the
differential equations of motion \cite{key-8-1}. However the relations
between both methods are easily found. It has been shown by M. Gasperini
and Veneziano {[}2,7{]} that the cosmological equations are rigorously
solved, in the isotropic case and for the vacuum ($T_{\mu\nu}=0=\sigma$)
by the solutions
\[
A(t)=A_{0}\left[\frac{t}{t_{0}}+\left(1+\frac{t^{2}}{t_{0}^{2}}\right)^{1/2}\right]^{1/\surd d}
\]

\par\end{flushleft}

\begin{flushright}
(2.2.1)
\par\end{flushright}

\begin{center}
and
\[
\bar{\phi}=-\frac{1}{2}\mathrm{ln}\left[\sqrt{V_{0}}t_{0}\left(1+\frac{t^{2}}{t_{0}^{2}}\right)\right]
\]

\par\end{center}

\begin{flushright}
Our equations (\ref{2.1.13}) are slightly different with the presence
of the factors $\frac{1}{a^{2,}}$, leading to the following conditions,
which must relate the scale factor to the sources properties:
\par\end{flushright}

\begin{center}
\[
\frac{1}{a^{2}}=-\frac{1}{3}e^{\bar{\phi}}\bar{\sigma}=-2e^{\bar{\phi}}\bar{p}=-\frac{2}{3}e^{\bar{\phi}}\bar{\rho}\Longrightarrow\bar{\rho}=3\bar{p},\;\bar{\sigma}=2\bar{\rho}
\]

\par\end{center}

\begin{flushright}
(2.2.2)
\par\end{flushright}

\begin{flushleft}
This shows that the solutions (2.2.2) are even valid in the presence
of matter, provided the conditions {[}2{]} $\gamma=1/3,$ $\gamma_{0}=2$
are satisfied, i.e. in the case of a pure radiation field :this is
an improvement over previous results.
\par\end{flushleft}

The curves representing $H(t)$ and $\bar{\phi}$ are similar to those
of {[}2{]} , (Fig. 4.7) with a bell-like shape for the curvature and
the dilaton kinetic energy. They show a pre-big bang inflationary
evolution, followed by a decelerated expansion. 

These solutions exhibit no possible acceleration of the universe expansion,
which is actually observed. This demonstrates that such an acceleration
is due to the presence of another kind of field, such as dark energy. 

The most recent experimental results on the observation of an acceleration
of the actual expansion of the universe is found in \cite{key-10-1}.
From the results published in this report, we can perform a fitting
of the continuous curve of $\dot{\tilde{A}}(t)$ represented on Fig.21
of \cite{key-10-1}. For that, we must first numerically solve the
first order differential equation which determine $\tilde{A}(t):$
\[
\dot{\tilde{A}}(t)\dot{=\tilde{A}(t)\times H_{0}(\Omega_{\Lambda}+\Omega_{M}(1+z)^{3}+(1-\Omega_{\Lambda}-\Omega_{M})(1+z^{2}))^{1/2}}
\]

\begin{flushright}
(2.2.3)
\par\end{flushright}

Performing this operation is not easy, because this is a stiff equation
which must be solved by efficient algorithms. This is done using the
Mathematica package \cite{key-8-1}. Concerning the variables, we
will use either $t$ or the redshift parameter $z$ , which taking
arbitrarily $\tilde{\tilde{A(0)}}=1,$ are related by
\[
1+z(t)=\frac{1}{y(t)},\; y(t)=\tilde{A}(t)
\]

\begin{flushright}
(2.2.4)
\par\end{flushright}

\begin{flushleft}
A numerical fitting of $z(t)$ Taking arbitrarily $\tilde{A}(0)=1$,
is:
\begin{eqnarray*}
1/\tilde{A}(t) & = & 1.0-70.62831490791527t+3740.638505154311t^{2}\\
 &  & -79103.96903009895t^{3}-1.578335812532445\times10^{7}t^{4}\\
 &  & -6.744184076495469`\times10^{9}t^{5}+8.136104416644515`\times10^{11}t^{6}\\
 &  & +1.1127413276575105\times10^{14}t^{7}-5.506299532617992\times10^{15}t^{8}\\
 &  & -8.075146184886697\times10^{17}t^{9}
\end{eqnarray*}

\par\end{flushleft}

\begin{flushright}
(2.2.5)
\par\end{flushright}

\begin{flushleft}
valid in the interval $-0.011\leqslant t\leqq0.008$, $t=0$ being
the present epoch. In fact the unit for $t$ is very large: relatively
to the present epoch, $\triangle z=0.5$ corresponds to $\triangle t=0.0066$s.
If we consider that the expansion of the universe began to accelerate
at $\triangle T=5My$ (cosmic time)\cite{key-9-1}, it follows $\triangle T/\triangle t=2.38\times10^{19}$.
So the time interval between $t_{1}=-0.011$ and $t_{2}=0.008$ is
$\triangle T=14.35My$ and covers the whole life of the universe.
\par\end{flushleft}

$\tilde{A}(t)$ is an increasing function of $t.$ The curve $\dot{\tilde{A}}(t)$
, when expressed in terms of the redshift parameter $z(t)$, reproduces
the continuous curve of Fig. 20 \cite{key-10-1}.

\begin{center}
The equation which determines $\bar{\phi}(t)$ is deduced from (\ref{2.1.12}):
\[
\ddot{\bar{\phi}}+\left(1-\frac{\gamma_{0}}{2}\right)\frac{1}{\gamma}H\dot{\bar{\phi}}-\frac{\partial V}{\partial\bar{\phi}}=dH^{2}+\left(1-\frac{\gamma_{0}}{2}\right)\frac{1}{\gamma}\dot{H}+\frac{1}{2a^{2}}\left[3-\frac{1}{2\gamma}\left(1-\frac{\gamma_{0}}{2}\right)\right]
\]

\par\end{center}

\begin{flushright}
(2.2.6)
\par\end{flushright}

Now, the solutions that we have adopted for $A(t),\;\bar{\phi}(t),\; V(\bar{\phi})$
are not rigorous solutions of (\ref{2.1.12}). To get a rigorous solution,
we must add a correction $\delta V$ to the potential to compensate
for the other components of the cosmic field, such that
\begin{eqnarray*}
-\frac{3}{2a^{2}}-\delta V=e^{\bar{\phi}}(\bar{\rho}-\bar{\rho}_{r})\\
-\frac{1}{4a^{2}}=\frac{1}{2}e^{\bar{\phi}}(\bar{p}-\bar{p}_{r})=\frac{1}{2}e^{\bar{\phi}}\gamma(\bar{\rho}-\bar{\rho_{r}})\\
-\frac{3}{2a^{2}}+\delta V-\frac{\partial}{\partial\bar{\phi}}(\delta V)=\frac{1}{2}e^{\bar{\phi}}(\bar{\sigma}-\bar{\sigma}_{r})
\end{eqnarray*}

\begin{flushright}
(2.2.7)
\par\end{flushright}

From which results:
\begin{eqnarray*}
-\frac{1}{a^{2}} & = & 2e^{\bar{\phi}}\gamma(\bar{\rho}-\bar{\rho}_{r})\\
\delta V & = & e^{\bar{\phi}}(3\gamma-1)(\bar{\rho}-\bar{\rho_{r}})\\
\frac{\partial}{\partial\bar{\phi}}(\delta V) & = & e^{\bar{\phi}}\left(6\gamma-1-\frac{\gamma_{0}}{2}\right)(\bar{\rho}-\bar{\rho}_{r})
\end{eqnarray*}

\begin{flushright}
(2.2.8)
\par\end{flushright}

It follows that Eqn (\ref{2.2.6}) is now
\begin{eqnarray*}
\ddot{\bar{\phi}}+\left(1-\frac{\gamma_{0}}{2}\right)\frac{1}{\gamma}H\dot{\bar{\phi}} & = & dH^{2}+\left(1-\frac{\gamma_{0}}{2}\right)\frac{1}{\gamma}\dot{H}-e^{\bar{\phi}}\frac{\gamma_{0}}{2}(\bar{\rho}-\bar{\rho}_{r})=dH^{2}+\left(1-\frac{\gamma_{0}}{2}\right)\frac{1}{\gamma}\dot{H}+\frac{1}{4a^{2}}\frac{\gamma_{0}}{\gamma}
\end{eqnarray*}
with

\begin{flushright}
\[
\gamma_{0}=2,\gamma=\frac{1}{3}
\]

\par\end{flushright}

\begin{flushright}
(2.2.9)
\par\end{flushright}

With these values for the two constant, Eqn (\ref{2.2.9}) takes the
remarkable simple form
\[
\ddot{\bar{\phi}}=dH^{2}+\frac{1}{4a^{2}}(\gamma_{0}/\gamma)
\]

\begin{flushright}
(2.2.10)
\par\end{flushright}

We have the following values for the density \cite{key-10-1}:
\[
\rho=\Omega_{M}(1+z)^{3}+\Omega_{\Lambda}+(1-\Omega_{\Lambda}-\Omega_{M})(1+z)^{2}
\]

A plot of the density as a function of $z$ is given in Fig. \ref{Fig1}.

\begin{center}
\textbf{}
\begin{figure}
\textbf{\protect\caption{The density $\rho$ as a function of $z$}
}
\end{figure}

\par\end{center}

We make a numerical integration of Eqn. (\ref{2.2.9}), we need to
evaluate $H(t)$ and $\dot{H}(t)$ which are given by the following
expressions: 
\[
H(t)=\frac{\dot{a}(t)}{a(t)},\;\dot{H}(t)=\left(\frac{\ddot{a}(t)}{a(t)}\right)-\left(\frac{\dot{a}(t)}{a(t)}\right)^{2}
\]
$1/\tilde{A}(t)$ being given by (\ref{CASS=0000C9 : R=0000E9f : 2.2.6}).

We have also:
\[
H(z)=69.9868+28.7795\times z+22.612\times z^{2}-2.45536\times z^{3}
\]

We obtain the curve given in Fig. \ref{Fig2}, showing the evolution
of $\bar{\phi}(t)$ with the initial condition $\bar{\phi}(t=-0.0066\mbox{s})=0$,
fixing the beginning of an accelerating expansion at $T=-5My.$ $\bar{\phi}(t)$,
correlating to the evolution of the expansion, is decreasing during
the decelerating expansion, and then increasing during the accelerating
expansion phase. 

\begin{center}
\textbf{}
\begin{figure}
\begin{centering}
\includegraphics[width=15cm]{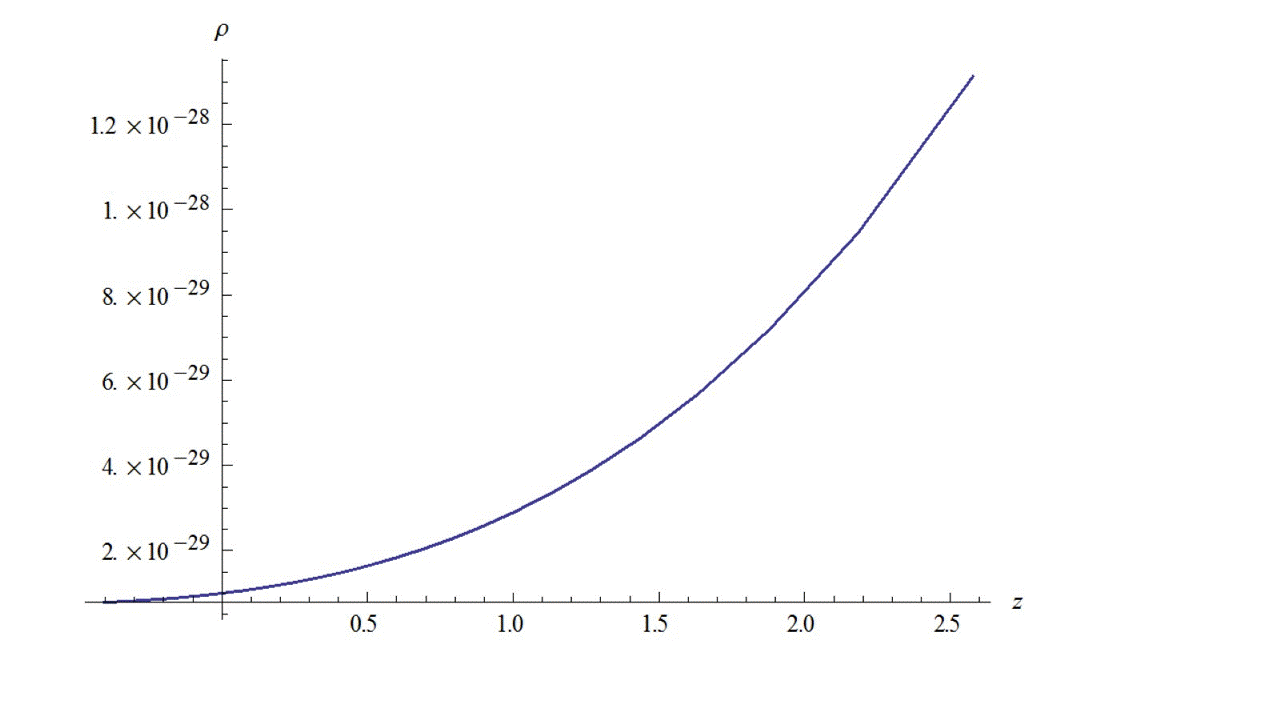}
\par\end{centering}

\textbf{\protect\caption{The dilaton $\bar{\phi}(t)$}
\label{Fig1}}
\end{figure}

\par\end{center}

\textit{Matching of the solutions: }To get a full description of the
universe evolution, we must first propagate Gasperini solution as
soon as the radiation field dominates,- as we have seen that the bell-shaped
GV-solution is still valid.\textit{ }For later time, we must use the
$\Lambda CDM$ solution , matching both at the junction point. 

This is done in the following way.

The GV-solution (Gasperini-Veneziano) is

\begin{center}
$H(t)=\left(d\times(t^{2}+t_{1}^{2})\right)^{-0.5},\;\dot{\bar{\phi}}(t)=-t/(t^{2}+t_{1}^{2})$
\par\end{center}

Matching $H(\Lambda CDM)$ and H(GV) at $t=-0.006$ fixes $t_{1}=0.000678617$.
It results that $\dot{\bar{\phi}}(t=-0.006)=166.667$. Taking $\bar{\phi}(-0.006)=0$
we get the curve given in Fig. \ref{Fig3}, showing the increasing
of $\bar{\phi}(t)$.

\textbf{\label{Fig2}}

\begin{center}
\textbf{}
\begin{figure}
\begin{centering}
\includegraphics[width=15cm]{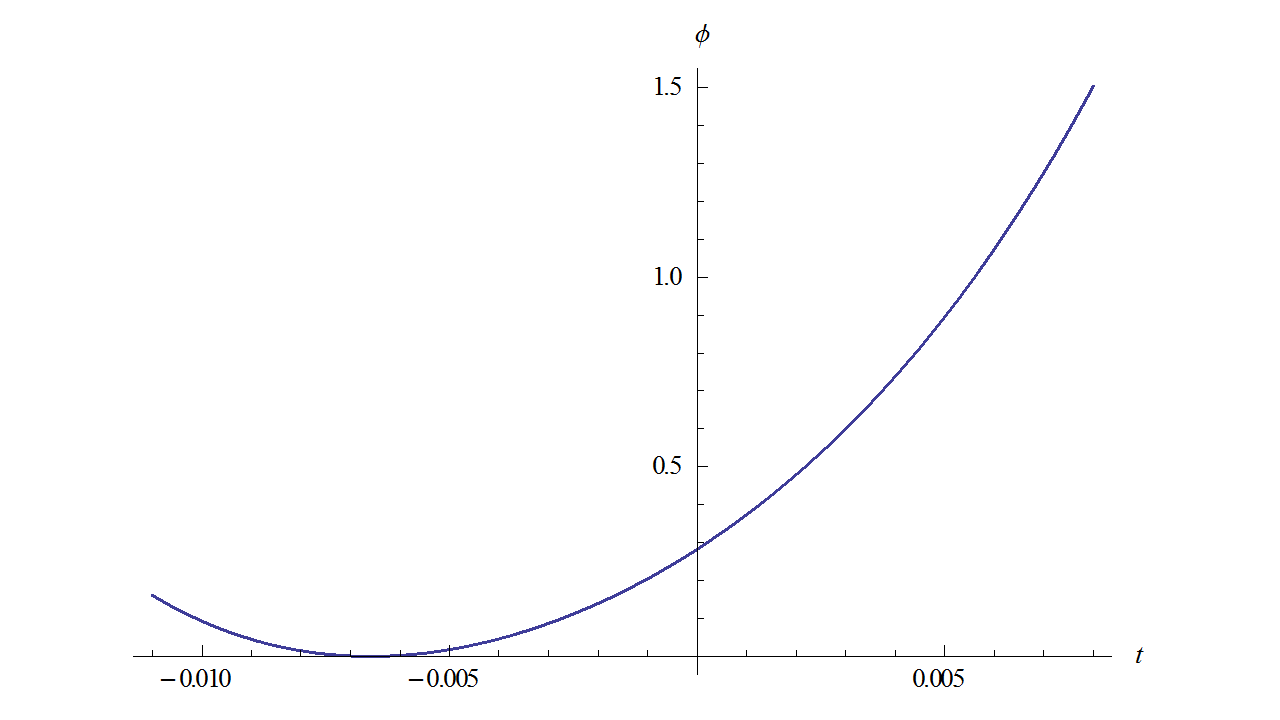}
\par\end{centering}

\textbf{\protect\caption{The global solution for the dilaton field $\bar{\phi}(t)$.}
\label{Fig3}}
\end{figure}

\par\end{center}

\section{Conclusion.}

In this article, I have developped models for the evolution of the
universe, using the Weiss-Zumino-Witten method on the $SU_{2}$ group.
Evolution equations for the fields variables have been established
in the simplest case where the graviton-dilaton system is described
by the dilaton field $\phi(t)$ . Solution are numerically computed,
taking care of the stiffness of the equations . These solutions describe
well the actual acceleration of the expansion which, according to
the most present measurements began around 5 billion years ago. So
this simple models show the ability of string theory, to describe
the actually observed evolution of the universe, giving an interpretation
in terms of the dilaton field.

\end{document}